\documentclass[aps, prd, nofootinbib, amsmath, preprintnumbers, superscriptaddress,secnumarabic, notitlepage,twocolumn]{revtex4-2} 
\usepackage{amssymb,amsmath,mathrsfs,graphicx,latexsym,amsthm,slashed,eso-pic,hyperref}
\usepackage{comment}
\usepackage{bbold}

 \def\d{{\rm d}}  
       
\newcommand{\beq}{\begin{equation}} \newcommand{\eeq}{\end{equation}}
\newcommand{\bea}{\begin{eqnarray}} \newcommand{\eea}{\end{eqnarray}}

\newcommand{\be}{\begin{equation}}
\newcommand{\ee}{\end{equation}}

\definecolor{orange}{rgb}{1,0.5,0}

\usepackage{slashed}
\usepackage{appendix}
\usepackage{siunitx}

\begin{document}
\begin{flushright}
KANAZAWA-23-07    
\end{flushright}

\title{Boosted dark matter from a phantom fluid}
\author{James M.\ Cline}
\email{jcline@physics.mcgill.ca}
\affiliation{McGill University, Department of Physics, 3600 University Street, Montr\'eal, QC H3A 2T8, Canada}
\author{Matteo Puel}
\email{matteo.puel@mail.mcgill.ca}
\affiliation{McGill University, Department of Physics, 3600 University Street, Montr\'eal, QC H3A 2T8, Canada}
\author{Takashi Toma}
\email{toma@staff.kanazawa-u.ac.jp}
\affiliation{Institute of Liberal Arts and Science, Kanazawa University, Kanazawa, Ishikawa 920-1192 Japan}
\affiliation{Institute for Theoretical Physics, Kanazawa University, Kanazawa, Ishikawa 920-1192 Japan}

\date{\today}

\begin{abstract}
It is known that theories of phantom
dark energy, considered as quantum fields, predict a continuous production of positive- plus negative-energy particles, from spontaneous decay of the vacuum.  We show that this can be a new source of boosted dark matter or radiation, with consequences for direct detection.  We set constraints on such models using data from the XENONnT experiment, and we show that recent excess events reported by the DAMIC experiment can be consistently described as coming from dark radiation, produced by vacuum decay, interacting with electrons.

\end{abstract}


\maketitle

{\bf Introduction.} Since the early determinations of the dark energy equation of state $w$ by the WMAP experiment \cite{WMAP:2003elm}, theorists have been exploring the possibility of getting $w<-1$ from scalar fields with wrong-sign kinetic terms, known as phantom models \cite{Caldwell:1999ew}.  These fields are almost invariably considered only at the classical level in the literature, but it was realized that when they are quantized, they inevitably lead to the spontaneous decay of the vacuum into negative-energy phantom (or ghost) particles plus positive-energy normal particles
\cite{Cline:2003gs}.  To avoid catastrophic production of gamma rays, an ultraviolet cutoff $\Lambda \lesssim$ several MeV must be imposed on the ghost momenta.

The positive energy particles produced by such vacuum decays would include any species with mass below $\Lambda$, including light dark matter particles.
They would have typical momenta of order
$\Lambda$, much larger than ordinary cold dark matter would have.  This provides a new source of {\it boosted}
dark matter \cite{Agashe:2014yua}, which 
has attracted attention by making light dark matter particles, which would normally carry too little energy to be observed, accessible to direct detection.

In a companion paper \cite{Cline:2023cwm} (hereafter denoted CPTW), we have shown that purely gravitationally coupled ghosts could not produce a cosmologically relevant density of 
the phantom fluid.  However if ghosts couple to normal matter more strongly than gravity, they can have an appreciable effect on the Hubble expansion and on cosmological perturbations.  For definiteness, we will follow CPTW by assuming a vector interaction 
\be
    {\cal L} = {i\over M^2}(\phi^*\!\!\stackrel{\leftrightarrow}{\partial}_\mu\!\phi)\,\bar\nu_s\gamma^\mu\nu_s,
    \label{ghost-nus-int}
\ee
coupling massless complex ghosts $\phi$ to
a ``sterile neutrino'' dark matter component with mass $m_{\nu_s}$.  

In this work we will not insist that $\nu_s$ comprises all of the dark matter, but rather consider it as an additional dark species.  In fact, the case where $m_{\nu_s}=0$  is of particular interest, since then the energy density of $\nu_s$
produced by vacuum decays is exactly canceled by that of the ghosts (hence the additional effective number of neutrinos remains $\Delta N_{\rm eff}=0$), and the number density of $\nu_s$ could be much greater than that of a 
conventional dark radiation species, carrying energies of order $\Lambda$.  

In CPTW it is shown that
cosmological data impose the limits 
\be
    {M\over {\rm GeV}} \gtrsim 
    \left\{\begin{array}{ll}
        4.6\times 10^8\, \hat\Lambda^{1.3-0.05\log_{10}\hat\Lambda},& m_{\nu_s}=0\\
        5.0\times 10^{13}\,\hat\Lambda^{2.25},& m_{\nu_s}\sim 0.5\Lambda
        \end{array}
            \right.\,,
    \label{cosmo-const}
\ee
depending on the value of $m_{\nu_s}$,
where $\hat\Lambda = \Lambda/{\rm MeV}$.
(For $0.01\,\Lambda \lesssim m_{\nu_s} \lesssim 0.5 \,\Lambda$
the limit is weaker than that for $0.5\,\Lambda$ by a factor of $\lesssim 3$.)  
For massless $\nu_s$, the momentum distribution function, including the effects of redshift, was found to be
\be
    p^2 f_{\nu_s}(p)\, dp = A \Lambda^4 H_0^{-1} x^2 e^{-\pi x^2} dx\,,
    \label{eq:nusdensity}
\ee
where $x=p/\Lambda$, $H_0$ is the Hubble constant, and $A = 3.9\times 10^{-5}(\Lambda/M)^4$.   It is normalized such that the number
density is given by $n_{\nu_s} = 
\int dp\, p^2 f_{\nu_s}(p) = A\Lambda^4/(4\pi\, H_0)$.
The factor of $H_0^{-1}$ reflects the fact that $\nu_s$ particles are produced at a constant rate (proportional to $M^{-4}$) throughout the history of the Universe.
Since the exotic particles are produced with large momentum $\sim \Lambda$, we will assume that they do not have time to cluster within galaxies, and they therefore maintain their cosmological densities for direct detection.

\bigskip
{\bf Dark matter-electron scattering.}
To make contact with direct detection experiments, we assume there is a similar interaction to Eq.\ (\ref{ghost-nus-int}) between the dark $\nu_s$ and electrons,
\be
    {1\over M'^2}(\bar e\gamma^\mu e)\,(\bar\nu_s\gamma_\mu\nu_s),
    \label{nu-e-int}
\ee
but with different strength $1/M'^2$.  It will turn out that the cosmological constraints (\ref{cosmo-const}) are too strong to give an observable direct signal unless $M' < M$.  The new interaction gives rise to a  squared matrix element for $\nu_s e\to\nu_s e$
scattering,
which at low energies is given by
\be
    \overline{\left|{\cal M}(q)\right|^2} \cong 4{(4 m_e^2 E_\nu^2- s\, q^2)\,\over M'^4},
    \label{M2eq}
\ee
where $E_\nu$ is the incoming $\nu_s$ energy, $s = m_e^2 + 2 m_e E_\nu + m_{\nu_s}^2$, and $q$ is the three-dimensional momentum transfer.  
The direct detection signal in the phantom fluid model is determined by just two combinations of parameters: $\Lambda$, which sets the scale of the energy spectrum, and $\sqrt{M M'}$ which governs the interaction rate of $\nu_s$ created by decay of the vacuum.
(Notice that $n_{\nu_s} \sigma_{e\nu_s}\sim (MM')^{-8}$.)  

\begin{figure*}[t]
\centerline{
\includegraphics[scale=0.306]{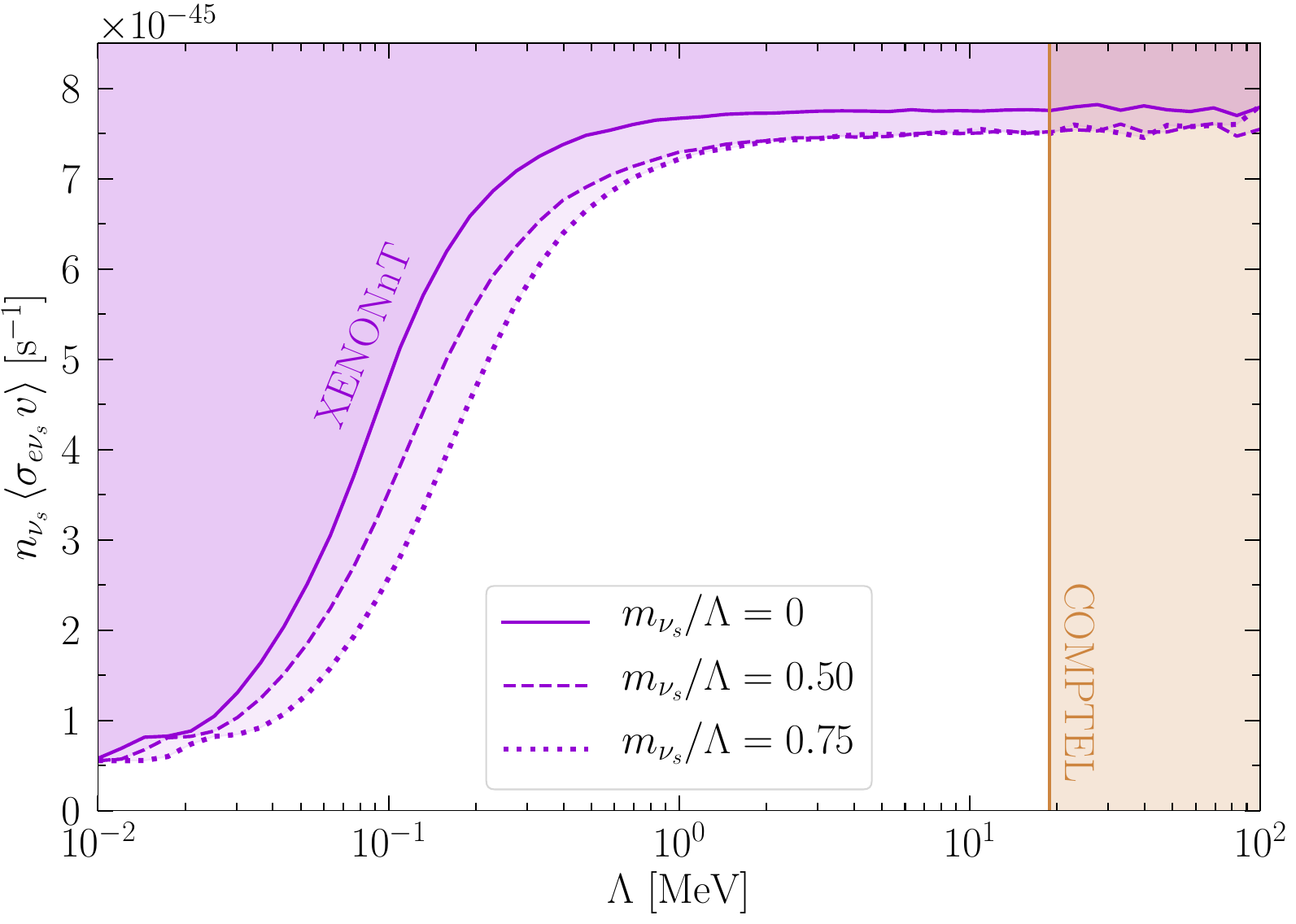}\hfil
\includegraphics[scale=0.3]{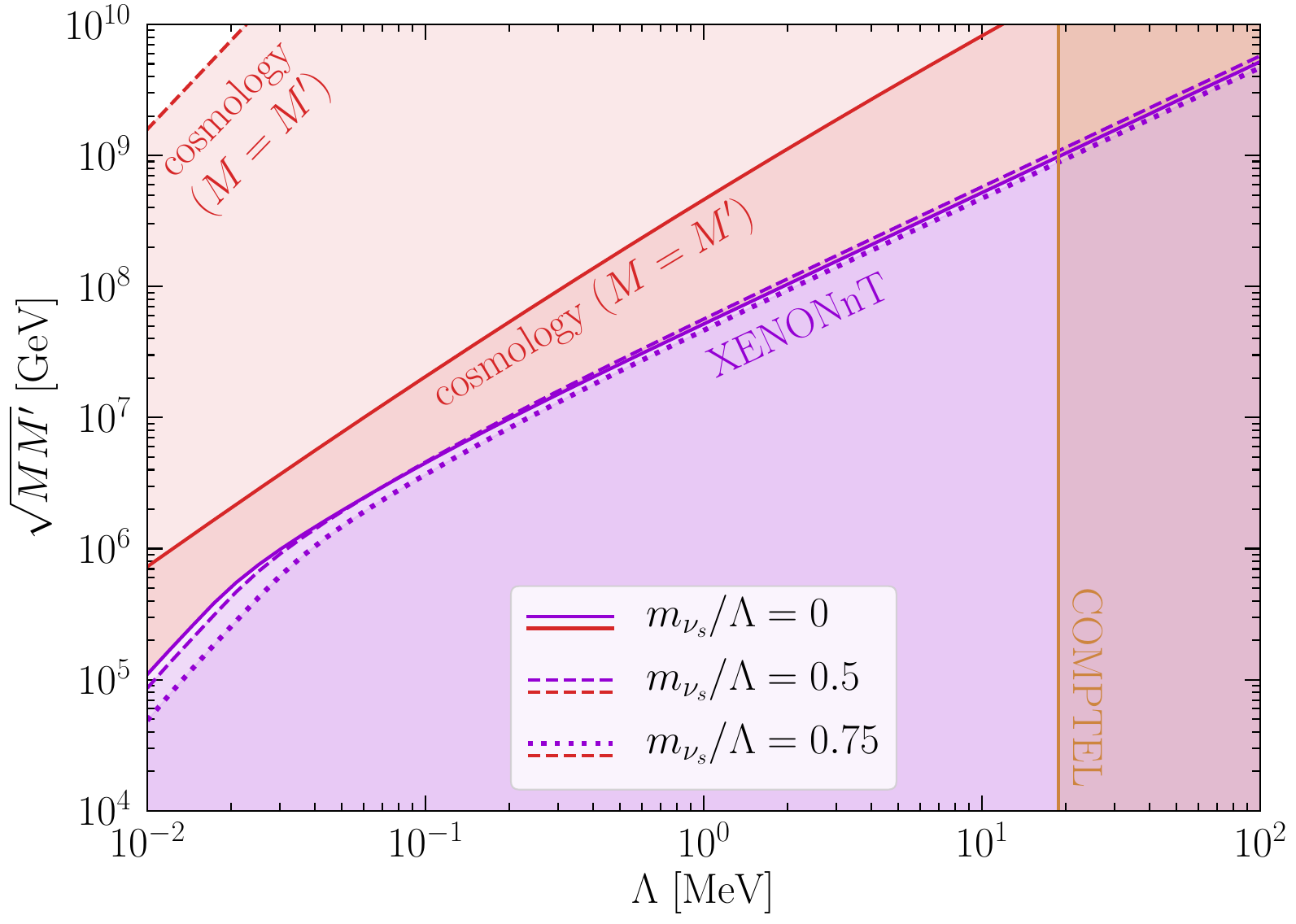}}
\caption{\textit{Left}: shaded regions excluded by XENONnT (violet) in the $\Lambda$ -- $n_{\nu_s}\langle\sigma_{e\nu_s}v\rangle$ plane, for $\nu_s$-$e$ scattering in the phantom fluid model.  Three different neutrino masses $m_{\nu_s} / \Lambda = (0,\,0.5,\, 0.75)$ are indicated.
The region labeled ``COMPTEL'' is ruled out by diffuse gamma-ray background 
\cite{Cline:2023cwm}.
\textit{Right}: XENONnT lower bounds on
$\sqrt{MM'}$ versus $\Lambda$ for 
the same three $m_{\nu_s}$ values
(violet curves).  The red curves show
the corresponding cosmological constraints from Ref.\ \cite{Cline:2023cwm} under the assumption that $M'=M$.
}
\label{fig:xenon-plots}
\end{figure*}

Interest in boosted dark matter was spurred by observations of excess electron recoil events by the XENON1T experiment
\cite{XENON:2020rca}, with energies
$\lesssim 3\,$keV.  New physics explanations involving boosted light particles were shown to fit the data
\cite{Kannike:2020agf,Fornal:2020npv}
(as well as models with nonboosted dark matter).  These explanations were later ruled out by stronger limits from the XENONnT experiment \cite{XENON:2022ltv}, suggesting that the XENON1T excess was due to a background from trace amounts of tritium.  We start by determining the constraints on the phantom fluid model from XENONnT.

\bigskip
{\bf Constraints from XENONnT.}
The differential cross section between an electron in the xenon atoms and $\nu_s$ at a fixed energy $E_{\nu}=\sqrt{p_{\nu}^2+m_{\nu_s}^2}$ is given by~\cite{Roberts:2015lga,Roberts:2016xfw,Roberts:2019chv}
\be
{\d\sigma \over dE_R} = 
{1\over 32\pi\, p_\nu^2\, m_e^3\,\alpha^2}
\int_{q_-}^{q_+} dq\,q\,
\overline{\left|{\cal M}(q)\right|^2}\,
K(E_R,q)\,,
\ee
where $K(E_R,q)$ is the atomic excitation factor, and the integration limits are
\be
q_{\pm}=p_{\nu}\pm\sqrt{p_{\nu}^2-2E_{\nu}E_R+E_R^2}\,.
\ee
We numerically fit the function
$K(E_R,q)$ from Fig.~7 in Ref.~\cite{Roberts:2019chv}, using the approximation
\be
K(E_R,q)=\left(\frac{E_R}{E_\mathrm{ref}}\right)^{-0.87}K_\mathrm{ref}\left(\frac{q}{\mathrm{MeV}}\sqrt{\frac{E_\mathrm{ref}}{E_R}}\right)\,,
\ee
where $K_\mathrm{ref}(q)$ is evaluated at a reference recoil energy $E_\mathrm{ref}=2~\mathrm{keV}$.\footnote{This procedure gives similar results to using accurate interpolation tables for $K(E,q)$ provided by Ref.\ \cite{Caddell:2023zsw}.  We thank B.\ Roberts for bringing it to our attention.}  The predicted differential event rate to compare to data requires averaging over the $\nu_s$ phase space distribution
(\ref{eq:nusdensity}),
\be
    {dR\over dE_R} = 
        n_T\, \epsilon_\mathrm{eff}(E_R)
            \int_{p_{\rm min}}^{2\Lambda} dp_\nu\,p_\nu^2
                \,f_{\nu_s}(p_\nu)\,
    {d\sigma v\over dE_R}\,.
\ee
where $v \cong p_\nu/E_\nu$ is the relative velocity, 
$n_T=4.2\times10^{27}~\mathrm{ton}^{-1}$ is the number density of xenon atoms and
$\epsilon(E_R)$ is the efficiency factor of the XENONnT detector~\cite{XENON:2022ltv}.
The minimum momentum required for recoil energy $E_R$ is 
$p_{\rm min} = (m_{\nu_s}+m_e)(E_R/2 m_e)^{1/2}$.

We have computed the expected values of $dR/dE_R$ in the energy bins corresponding to the XENONnT experimental observations, and constructed the $\chi^2$ function with respect to their background model and observed values.  The resulting 95\% confidence level limits  are presented in Fig.\ \ref{fig:xenon-plots}.  The left plot shows the
maximum allowed value of $\nu_s$ density times cross section, $n_{\nu_s}\langle\sigma_{e \nu_s}v\rangle$, versus the phantom cutoff-scale $\Lambda$.  The
right plot gives the corresponding lower limit
on the fundamental physics parameter $\sqrt{MM'}$.
For comparison, we also show the stronger cosmology bounds (\ref{cosmo-const}) in the case where $M'=M$, illustrating the need for $M'<M$ to get an observable direct signal.

\bigskip
{\bf DAMIC excess events.} Recently, the DAMIC experiment at SNOLAB reported excess events at recoil energies $\lesssim 0.4\,$keV  \cite{DAMIC:2023ela}.  We interpret them as being due to $\nu_s$-$e$ scattering like for XENON,
focusing on the massless case $m_{\nu_s}=0$ in which cosmological constraints on $\sqrt{MM'}$ are weaker, allowing us to probe higher values of the new-physics scale $M'$.

The DAMIC detector uses silicon charged-current device (CCD) technology, and similarly to xenon, requires understanding of bound-state properties of electrons in the atoms to analyze very low-energy scattering.  However for all but the lowest energy bin (of width 50\,eV), the electron energies well exceed the Si ionization energy, and we can approximate the electrons as being free.
(We will come back to the lowest bin below.)  

\begin{figure}[t]
\begin{center}
\includegraphics[scale=0.4]{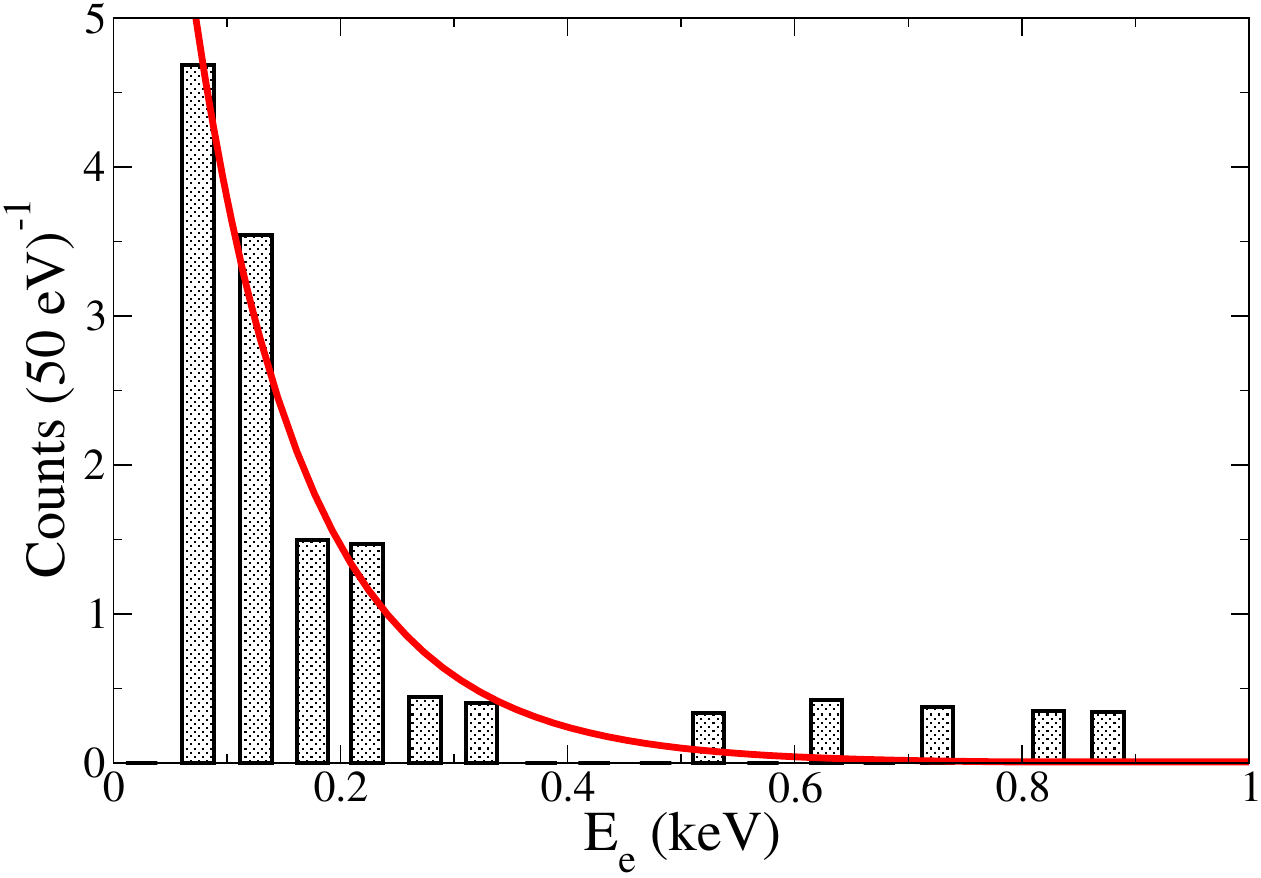}
\caption{DAMIC excess events (bars) after subtracting expected background, versus prediction from boosted $\nu_s$-electron scattering (red) with $\Lambda = 10$\,keV cutoff
momentum.
}
\label{fig:damic}
\end{center} 
\end{figure}

The differential cross section times relative velocity of $\nu_s$ scattering on free electrons in our model is 
\be
    {d(\sigma v)\over dE_e} = {m_e\over 2\pi M'^4}\,
    \left(1 - {m_e E_e\over 2p^2}
    \right)\Theta\left(1 - {m_e E_e\over 2p^2}
    \right)
    \,,
\ee
where $\Theta$ is the Heaviside function.  The average scattering rate per electron is found by integrating over the $\nu_s$
phase space density,
${dR/dE_e} =\int dp\,p^2\, f_{\nu_s}(p)\, {d(\sigma v)/dE_e}$.

By comparing the prediction for $dR/dE_e$ to the 
observed excess spectrum, after subtracting expected background contributions, as shown in Fig.\ \ref{fig:damic}, we find that the shape is best fit for $\Lambda \cong 10.0\,$keV, which is consistent with the diffuse x-ray bound.    Numerically, the total rate per electron is given by $R=1.8\times 10^{-6}\Lambda^{10}/(M^4 M'^4 H_0)$,
while the measured excess corresponds to 7 events per kg-day,
or $R = 2\times 10^{-52}$\,MeV per electron,
in natural units.  Equating the two, we find that 
\be
    (MM')^{1/2} \cong 1.3\times 10^{5}\,{\rm GeV}
\label{damic-norm}
\ee
to obtain the observed DAMIC excess.  This violates (by a factor of $\sim 5$) the cosmological constraint shown in the right panel of Fig.\ \ref{fig:xenon-plots}, under the assumption $M'\sim M$.   If the two interactions occur at different scales such that $M'\ll M$, this constraint can be satisfied.  For example if
$M=8.4\times 10^5$\,GeV, saturating the cosmological bound, then $M' \cong 19\,$TeV
is needed to match the DAMIC excess.  Significantly, the
required value (\ref{damic-norm}) is consistent with the XENONnT
constraint we derived above, which gives $(MM')^{1/2} > 1.0\times 10^5\,{\rm GeV}$.

To demonstrate that the lowest energy bin has a much smaller signal, we rederived the rate in the case of relativistic dark particle collisions, using appendix A of Ref.\ \cite{Essig:2015cda}.  The main differences arise from the energy-conserving delta function, which takes the form $\delta(E_e + |\vec p - \vec q| - p)$, the fact that $p^2$ appears in the matrix element (\ref{M2eq}), and
 the distribution function (\ref{eq:nusdensity}) depends on momentum $p$ rather than velocity.
 This results in the replacements
 \be
    {\rho_\chi\bar \sigma_e\over 2 m_\chi \mu^2_{\chi e}}\to {1\over M'^4},\quad
        \eta \to 2\pi^2\int_{p_{\rm min}}^{2\Lambda}
        dp\, p^2 f_{\nu_s}(p)
\ee
in the formulas of Ref.\ \cite{Essig:2015cda},
where $p_{\rm min} = (q+E_e)/2$.  Although the dimensions of each change do not match, their product does.  It is then possible to read off the differential rate for a single crystal cell
from Eq.~(A.32) of Ref.\ \cite{Essig:2015cda}.  Averaging over the first bin of width $\Delta E_e$, it gives
\be
    \left\langle{dR\over dE_e}\right\rangle = 
    {2\alpha m_e^2\over M'^4\Delta E_e}
    \int_0^{\Delta E_e}dE_e\int {dq\over q^2}
        \,\eta\, |f_c(q,E_e)|^2\,,
\ee
where $f_c$ is the silicon crystal form factor that can be downloaded \cite{QED}.  Carrying out the integral, we find that $\langle dR/dE_e\rangle$ in the first bin is suppressed by a factor $\sim 10^{-3}$ relative to 
its maximum value, occurring at the second bin.
Since $f_c$ has not yet been computed for energies
$E_e>\Delta E_e=50\,$eV, it is not possible to do a more quantitative study of the transition between bound and free electrons at this time.

The scenario we presented is capable of explaining the DAMIC excess with a far smaller cross section,
\be
    \sigma_{e\nu_s} = {\langle p_\nu^2\rangle\over \pi M'^4} \sim 4.5 \times 10^{-56}\,\,{\rm cm^2}\,,
\ee
than would be possible with a conventional dark species.  The energy (and number) density of the massless $\nu_s$ fluid can be enormous, $\Omega_{\nu_s} = \rho_{\nu_s}/\rho_{\rm crit} \sim 10^{14}$ (as derived in CPTW for $\Lambda = 10$ keV), where $\rho_{\rm crit}$ is the critical density.
This is possible because the negative energy of the phantom particles exactly cancels $\rho_{\nu_s}$ when $m_{\nu_s}=0$.  
Despite such a high cosmological density of $\nu_s$,  at the time of recombination, when $n_{\nu_s}$ was smaller by a factor of $t_{\rm rec}/t_0 = 2\times 10^{-5}$ relative to today, the rate of
scatterings per electron was only $10^{-36}$/s, which would have a negligible impact on recombination in the early Universe.

\bigskip
{\bf Conclusions.}  We have presented a new cosmological mechanism for producing boosted dark matter or radiation, from the spontaneous decay of the vacuum in theories of phantom dark energy.  It can work for light dark matter, with mass below the ultraviolet cutoff $\Lambda \lesssim 18\,$MeV, above which the effective phantom description must revert to a normal theory with positive-energy particles.   The spectrum of boosted dark particle momenta is nonthermal, Eq.\ (\ref{eq:nusdensity}), and could in principle be distinguished from other mechanisms for exciting DM at late times.
We found that the model can give a good fit to recently observed excess events in the DAMIC experiment, with just two combinations of parameters:
$\Lambda$ and the product of the new-physics energy scales that determine the couplings of the DM to
the ghosts and electrons, respectively.

To identify other processes that could give complementary constraints on the $\nu_s$-$e$ interaction, it would be necessary to find astrophysical systems colder than
$\Lambda \sim 10\,$keV, that are sensitive to electrons being heated by scatterings with $\nu_s$.  White dwarf
stars have been used to constrain dark matter accumulation by its effect on their cooling \cite{Dasgupta:2019juq}.  We find that the phantom model has a negligible impact on the evolution of white dwarfs.  The rate of heating per electron is of order
$10^{-14}\,$keV/Gyr, while the core temperatures of white dwarfs cool at a rate of $\sim 1$\,keV/Gyr.

Another interesting aspect of phantom fluids coupling to visible matter is that the cutoff $\Lambda$ applies to the phantom momentum in a preferred reference frame \cite{Cline:2003gs}, that we have for simplicity assumed to coincide with the CMB rest frame.  It is therefore a Lorentz-violating construct, and will introduce Lorentz-breaking effects into the standard model through loops of ghosts and $\nu_s$.  It could also possibly lead to ``Axis of Evil'' \cite{Copi:2005ff} type effects in cosmology in the case that the two frames do not coincide. We defer such  studies to the future.

\bigskip
{\bf Acknowledgments.}  We thank
G. Alonso-Alvarez,
R.\ Essig, B.\ Roberts, 
A.\ Singal and T.-T. Yu
for helpful discussions or correspondence.  JC and MP are supported by the Natural Sciences and Engineering Research Council (NSERC) of Canada. TT was supported by JSPS KAKENHI Grant Number JP23H04004.

\bibliography{references}
\bibliographystyle{utphys}

\end{document}